\documentclass[%
reprint,
superscriptaddress,
showpacs,
preprintnumbers,
 amsmath,
 amssymb,
aps,
pre,
floatfix,
]{revtex4-1}

\usepackage{graphicx}
\usepackage[colorlinks,allcolors=blue]{hyperref}
\usepackage{amsmath,amssymb,amsthm}
\usepackage{array, booktabs, makecell}
\usepackage[utf8]{inputenc}
\usepackage[english]{babel}
\usepackage{blindtext}

\begin{document}

\title{Estimating physical properties from liquid crystal textures via machine learning and complexity-entropy methods}

\author{H. Y. D. Sigaki}
\author{R. F. de Souza}
\affiliation{Departamento de F\'isica, Universidade Estadual de Maring\'a, Maring\'a, PR 87020-900, Brazil}
\author{R. T. de Souza}
\author{R. S. Zola}\email{rzola@utfpr.edu.br}
\affiliation{Departamento de F\'isica, Universidade Estadual de Maring\'a, Maring\'a, PR 87020-900, Brazil}
\affiliation{Departamento de F\'{\i}sica, Universidade Tecnol\'ogica Federal do Paran\'a, Apucarana, PR 86812-460, Brazil}
\author{H. V. Ribeiro}\email{hvr@dfi.uem.br}
\affiliation{Departamento de F\'isica, Universidade Estadual de Maring\'a, Maring\'a, PR 87020-900, Brazil}

\begin{abstract}
Imaging techniques are essential tools for inquiring a number of properties from different materials. Liquid crystals are often investigated via optical and image processing methods. In spite of that, considerably less attention has been paid to the problem of extracting physical properties of liquid crystals directly from textures images of these materials. Here we present an approach that combines two physics-inspired image quantifiers (permutation entropy and statistical complexity) with machine learning techniques for extracting physical properties of nematic and cholesteric liquid crystals directly from their textures images. We demonstrate the usefulness and accuracy of our approach in a series of applications involving simulated and experimental textures, in which physical properties of these materials (namely: average order parameter, sample temperature, and cholesteric pitch length) are predicted with significant precision. Finally, we believe our approach can be useful in more complex liquid crystal experiments as well as for probing physical properties of other materials that are investigated via imaging techniques.
\end{abstract}
\maketitle

\section{Introduction}

Optical imaging techniques are important tools extensively used for probing a number of materials properties~\cite{opte}. These imaging techniques are non-destructive and particularly convenient for dealing with biological and other complex materials~\cite{plant}. Liquid crystals are among these materials widely studied via optical and image processing methods~\cite{degennes}. This occurs because liquid crystals are birefringent materials, and as such, simple polarized optical microscope imaging already access some of their important properties, including birefringence and sample thickness~\cite{prl}. Moreover, this technique estimates the local ordering properties (for instance, the director distribution) across a sample when coupled with variable retarders and different algorithms for fast and sensitive measurements~\cite{first}. This approach is known as LC-PolScope~\cite{pol} and has been used for fine imaging of defect cores in lyotropic liquid crystals~\cite{oleg1} and can describe the orientational order of active nematics~\cite{active}.

Despite the extensive use of optical imaging approaches in the study of liquid crystals~\cite{montrucchio1998,sastry20121,sastry20122,sastry20123}, much less attention has been paid to the problem of extracting physical parameters directly from images of these materials. This is an important issue since several physical parameters of liquid crystals are only obtained by adjusting theoretical models to cumbersome and time demanding experimental results. Examples include the microscopic order parameter, from which several other parameters characterizing the nematic phase are dependent~\cite{degennes}, and the pitch length of cholesteric liquid crystals. The latter is easily obtained under an optical microscope when the helical axis lies perpendicular to the viewing direction~\cite{zigzag}, but cannot be estimated from the most commonly used experimental arrangements, where the helix orients parallel to the viewing direction (often called Grandjean texture, used in reflective displays)~\cite{jap}.

In this context, image-based characterization of liquid crystals can benefit hugely from state-of-the-art machine learning techniques~\cite{hastie2013elements}. These approaches have been available since the 1990s, but it was only during the last decade that such methods gained impressive popularity in several areas of science where unveiling meaningful patterns in data is fundamental. Naturally, physics is not an exception, and indeed there are several recent works employing machine learning algorithms for studying many physical systems~\cite{arsenault2014machine,kusne2014fly,PhysRevLett.114.108001,kalinin2015big,carrasquilla2017machine,PhysRevE.95.032504,PhysRevE.95.062122,PhysRevE.96.011301,PhysRevLett.120.257204}. Here we present an approach that is capable of extracting physical properties of nematic and cholesteric liquid crystals directly from their textures images. Our approach is based on the evaluation of two complexity measures related to the arrangement of pixels in the textures, combined with simple machine learning algorithms employed for classification and regression tasks. We demonstrate the potential of this approach in a series of applications based on numerically-generated and experimental textures, from which physical parameters of liquid crystals are predicted with high accuracy.

In what follows, we present our results on simulated and experimental nematic liquid crystals, where we show that our approach predicts the average order parameter from simulated textures with an accuracy of $\approx$99\% and the sample temperature with an accuracy of $\approx$93\%. We further show that this approach classifies different pitches of cholesteric textures with an overall accuracy of $\approx$85\%. Next, we present our conclusions and methodological details of the proposed approach as well as our experimental procedure and numerical methods used for obtaining liquid crystal textures.  

\section{Results}

In all studies presented here, we have calculated two simple complexity measures directly obtained from liquid crystal textures: the permutation entropy $H$ and the statistical complexity $C$~\cite{BandtPompe2002,RossoLarrondoMartinPlastinoFuentes2007,Ribeiro_etal2012, Zunino2016679}. As detailed in Appendix~\ref{app:plane}, these two quantities are estimated from an ordinal probability distribution $P=\{p_1,p_2,\dots,p_n\}$, whose components represent the probability of finding a given two-dimensional local ordering pattern of size $d_x \times d_y$ (the embedding dimensions) over the pixels of a texture. The entropy $H$ quantifies the degree of disorder regarding the occurrence of these local patterns. A texture characterized by $H\approx1$ has pixels randomly distributed in space, while for $H\approx0$, these pixels appear practically at the same order along the texture. The complexity $C$ quantifies the degree of ``structuredness'' in the arrangement of pixels in a texture. A value of $C\approx0$ occurs for both extremes of order and disorder, whereas $C>0$ represents a texture with more complex spatial patterns.

We employ only these two measures as predictive features of physical properties of liquid crystals. The predictions are obtained by training the nearest neighbors algorithm~\cite{hastie2013elements} with the values of $H$ and $C$ from a set of textures in which the physical property in question is known, and next the trained algorithm is exposed to values of $H$ and $C$ from another set of textures in order to predict the physical property. The nearest neighbors algorithm is one of the simplest machine learning algorithms for classification and regression tasks~\cite{hastie2013elements}, but it already provides excellent accuracy in our results. Details of the implementation of this algorithm are given in Appendix~\ref{app:machinelearning}.

\subsection{Monte Carlo simulated textures}\label{res:montecarlo}
\begin{figure}[!ht]
\centering
\includegraphics[width=1\linewidth]{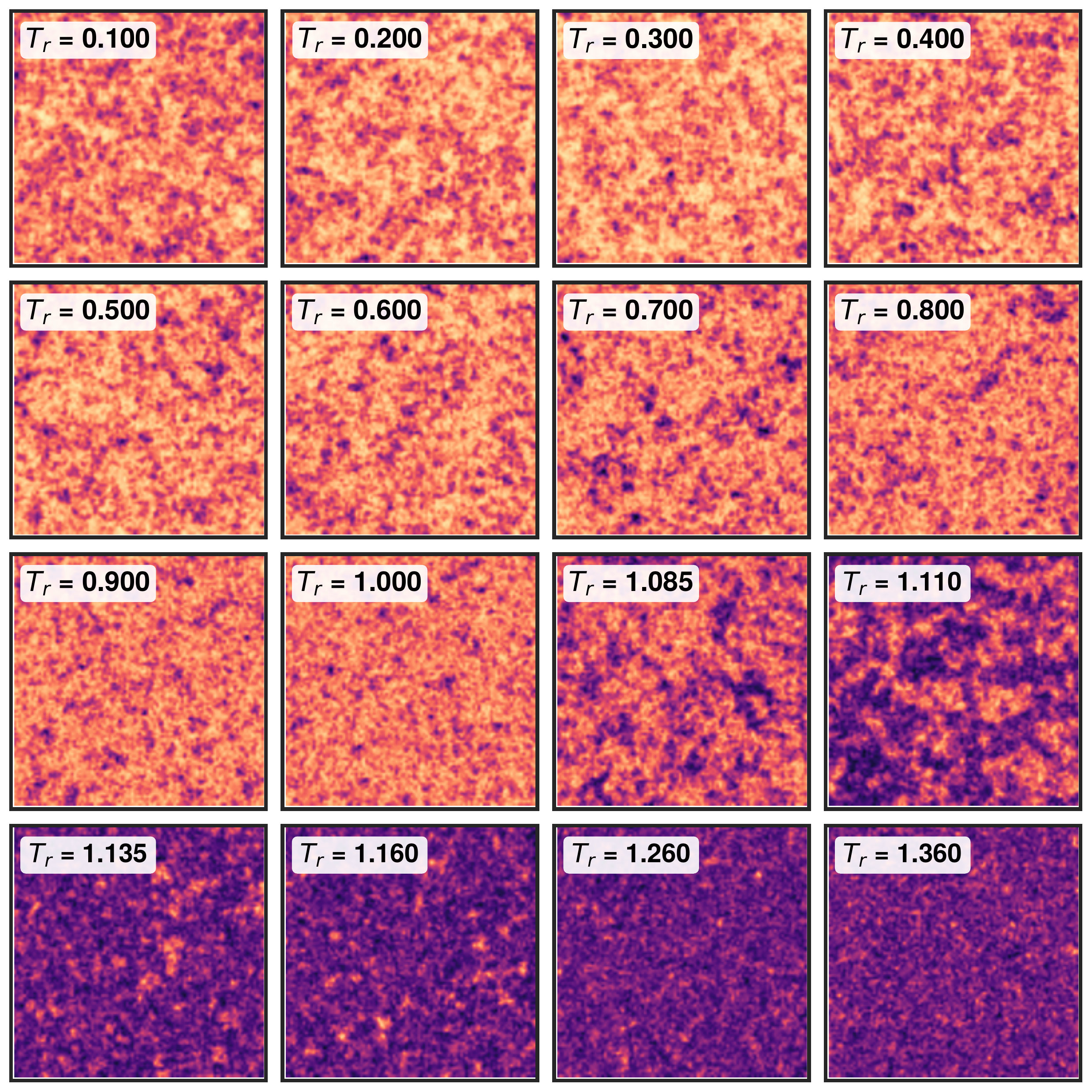}
\caption{{Examples of a nematic liquid crystal texture at different temperatures and phases.} These textures are generated by using the Monte Carlo method described in Appendix~\ref{app:montecarlo}. Each texture corresponds to a different reduced temperature $T_r$ (as indicated within the plots) and this system presents a nematic to isotropic phase transition at the critical temperature $T_c=1.1075$. We note that textures from different phases are easily distinguishable, while nematic textures ($T_r<T_c$) from different reduced temperatures are very similar to each other as well as the ones from the isotropic phase ($T_r>T_c$).}
\label{fig:1}
\end{figure}

We start by analyzing nematic textures generated by Monte Carlo simulations of the model described in Appendix~\ref{app:montecarlo}. Examples of these textures are shown in Figure~\ref{fig:1} for different reduced temperatures $T_r$. This system undergoes a nematic to isotropic phase transition when $T_r$ exceeds the critical temperature $T_c = 1.1075$. Each texture has a different average order parameter $p$ that depends on the temperature $T_r$ (see Eq.~\ref{eq:order_parameter} for details), and our goal is to predict the value of $p$ directly from these images by using the values of $H$ and $C$. We note that the textures exhibit visually distinct patterns between the nematic and isotropic phases; in particular, for $T_r>T_c$ we observe the emergence of isotropic domains that predominate in the texture as the temperature increases beyond $T_c$. While nematic and isotropic textures are easily distinguished from each other, even a well-trained eye of an experimental physicist will be in trouble for distinguishing among nematic textures with different temperatures (for instance, between $T_r=0.2$ and $T_r=0.6$) as well as among isotropic textures at different temperatures. 

\begin{figure}[!ht]
\centering
\includegraphics[width=1\linewidth]{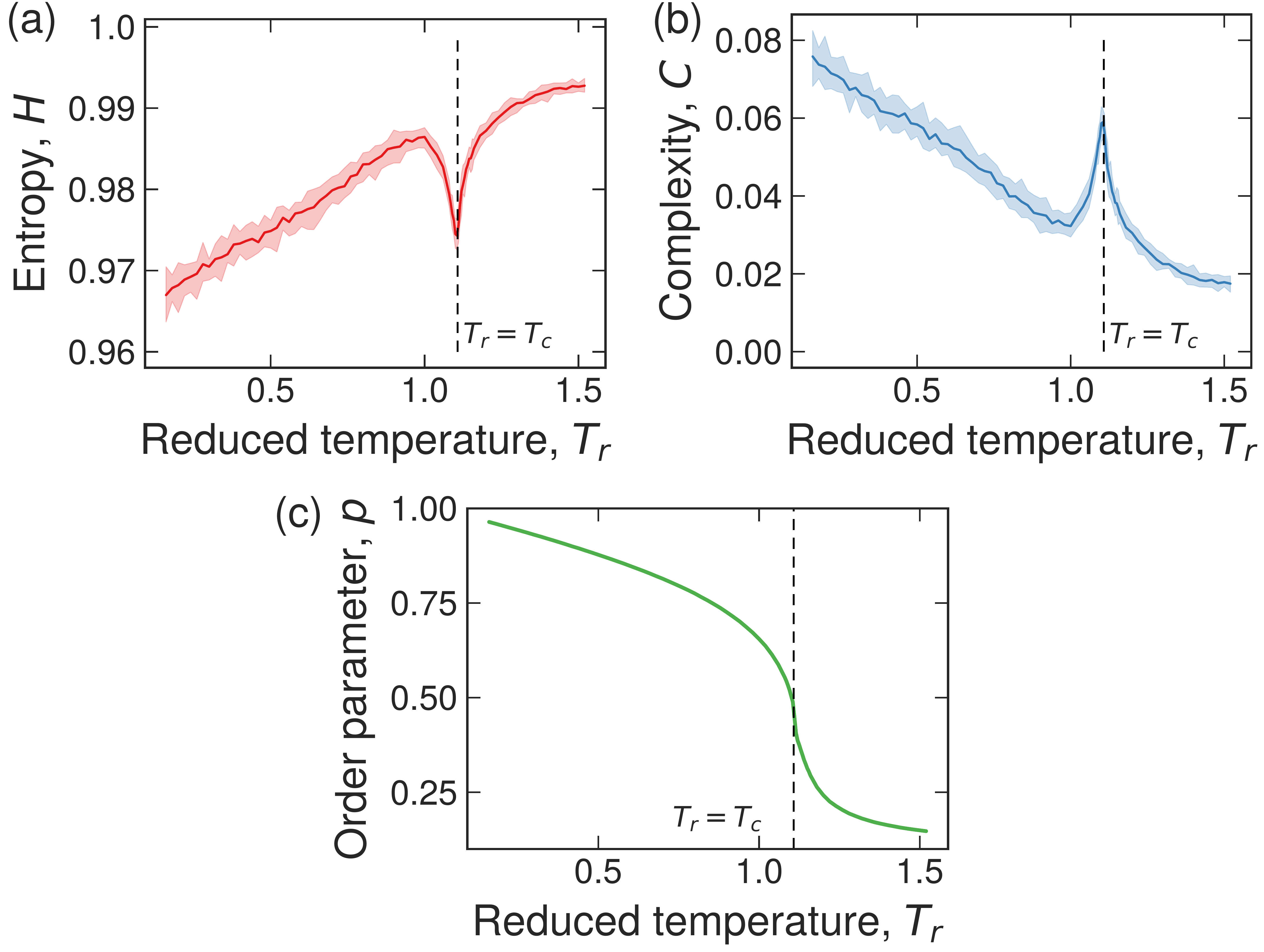}
\caption{{Dependence of the image quantifiers and the order parameter on the temperature.} (a) Values of the permutation entropy $H$ and (b) the statistical complexity $C$ as a function of the reduced temperature $T_r$. The solid curves represent average values over 50 realizations, and the shaded areas are the $95\%$ bootstrap confidence intervals. (c) Dependence of the order parameter $p$ versus the reduced temperature $T_r$. In all panels, the dashed line indicates the critical temperature ($T_c= 1.1075$) of the nematic-isotropic phase transition. We note that the phase transition is clearly and properly identified by the extreme values of the complexity measures.}
\label{fig:2}
\end{figure}

Despite the visual similarity among the textures, we show that the values of $H$ and $C$ are capable of distinguishing among these images as well as identifying the nematic-isotropic transition. To do so, we create a dataset composed of several realizations of simulated nematic textures for different temperatures $T_r$, and evaluate the values of $H$ and $C$ for each one. Figures~\ref{fig:2}A and \ref{fig:2}B show the dependence of the average values of $H$ and $C$ on the reduced temperature $T_r$. We observe that $H$ has a trend to increase with the temperature but shows a sharp minimum at $T_r=T_c$. Similarly, the values of $C$ tend to decrease with the temperature and present a sharp maximum at $T_r=T_c$.  Figure~\ref{fig:2}C depicts the dependence of the order parameter $p$ on the reduced temperature $T_r$, where the critical temperature $T_c=1.1075$ is defined as the value that maximizes the absolute value of the derivative of $p$ with respect to $T_r$. 

The well-defined dependence of the average values of $H$ and $C$ on the temperature $T_r$, combined with the fact that the order parameter $p$ is also a function of $T_r$, indicates that we can predict the values of $p$ directly from the images. It is worth noting that the values of $H$ and $C$ are not uniquely defined for a given temperature, displaying some random fluctuations associated with the process that generates these textures. Thus, to test the predictive power of these image quantifiers in a more practical situation, we have trained a $k$-nearest neighbors algorithm for the regression task of predicting the order parameter $p$ based on the values of $H$ and $C$ and a dummy variable that is zero for $T_c-0.05T_c\leq T_r\leq T_c$, 1 for $T_r>T_c$, and -1 for $T_r<T_c-0.05T_c$. This dummy variable is necessary due to the non-biunivocal relations between the image quantifiers and the temperature. However, the value of $T_c$ can be directly estimated from the image quantifiers. As detailed in Appendix~\ref{app:machinelearning}, the $k$-nearest neighbors algorithm is among the most straightforward statistical learning approaches that assigns to an unlabeled object (the value of $p$ from a texture) the most frequent label among the $k$ training nearest neighbors (the parameter of the method) in the features space (the extracted image features).

\begin{figure}[!ht]
\centering
\includegraphics[width=1\linewidth]{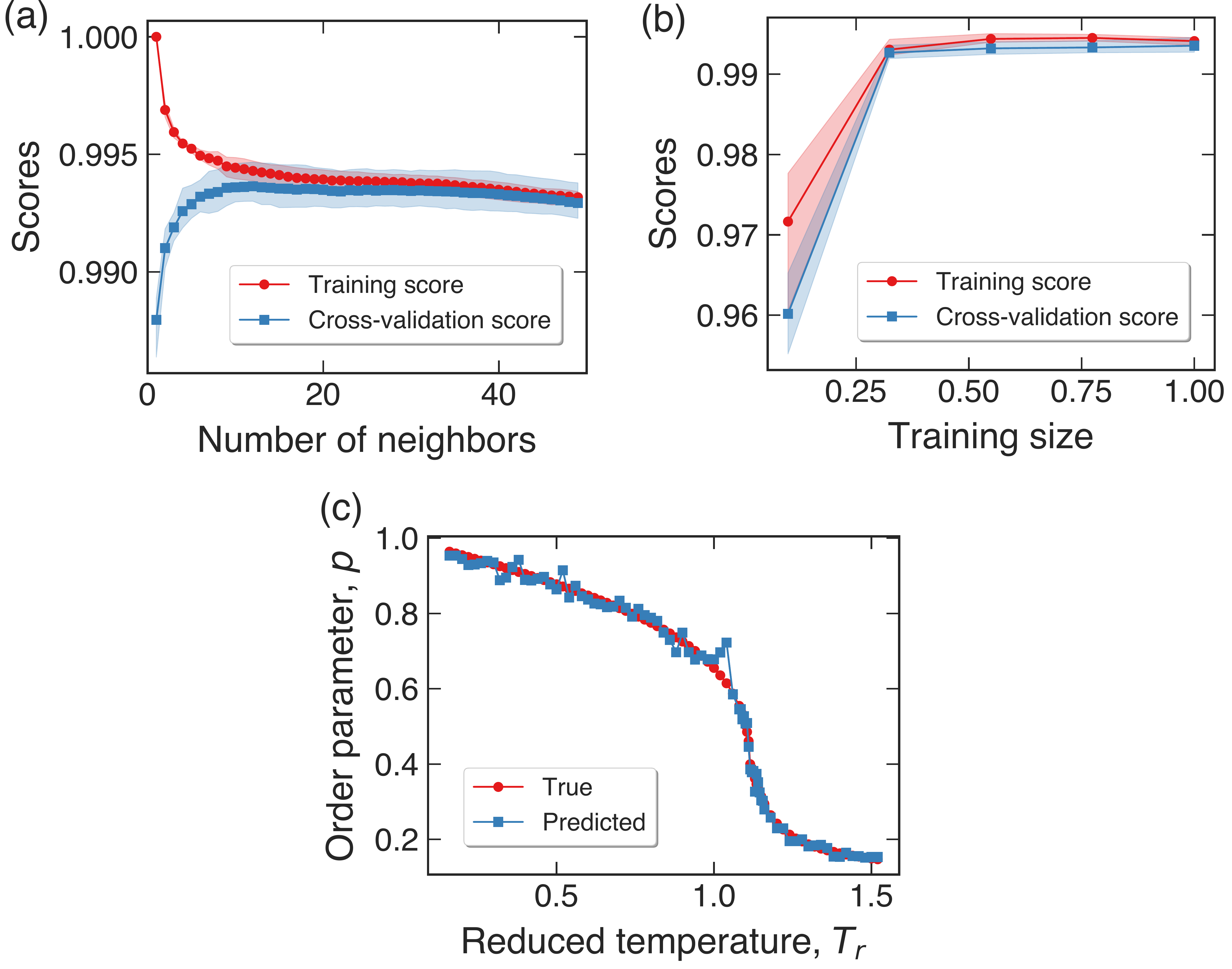}
\caption{{Predicting the order parameter $p$ with a machine learning algorithm.} (a) Training and cross-validation scores of the $k$-nearest neighbors algorithm as a function of the number of neighbors $k$. The scores represent the coefficient of determination ($R^2$) of the relationship between the predicted and true values, and can be interpreted as the percentage (or fraction) of data variation that is explained by the model. We note that the algorithm overfits the data when $k<10$, whereas for $k>20$ it starts underfitting the data. (b) Learning curves, that is, training and cross-validation scores as a function of the training size (fraction of the whole data used for training the model) with $k=15$. We observe no significant improvement in the cross-validation score when more than $35\%$ of the data is used. The shaded areas in both plots represent the $95\%$ confidence intervals obtained in a 3-fold-cross-validation splitting strategy. We note the high accuracy achieved by the algorithm ($\approx 99.2\%$). (c) True and predicted order parameter $p$ as a function of the reduced temperature $T_r$. These predictions were generated by exposing the trained regressor (with $k=15$) to a set of textures never presented before to the algorithm.}
\label{fig:3}
\end{figure}

Figure~\ref{fig:3}A shows the validation curves, that is, the training and cross-validation scores as a function of the number of neighbors $k$. We note that the algorithm overfits the data for $k<10$, that is, the algorithm is not complex enough to capture the underlying structure of the data. On the other hand, the algorithm begins to underfit the data for $k>20$, meaning that the learning method is getting too complex and modeling even the random noise in the training set. Figure~\ref{fig:3}B shows the learning curves, which represents the scores as a function of the fraction of the data used for training the algorithm (with $k=15$). We observe no significant improvement in the cross-validation score when more than 35\% of the data is used as training set. Thus, the $k$-nearest neighbors algorithm achieves a remarkable accuracy of $\approx 99.2\%$ in the regression task of predicting the order parameter $p$ solely based on the values of $H$ and $C$. We further observe that pratically the same accuracy is obtained when using only the values of $H$ or only the values of $C$. This happens because the values of $H$ and $C$ are strongly correlated to each other for these textures. However, in general, the values of $C$ are not a trivial function of $H$ and usually care additional information related to the ``structural'' complexity of images~\cite{Ribeiro_etal2012,Zunino2016679,sigaki2018history}. The performance of this algorithm is much higher than those obtained from simple baseline regressors that always predict the expected values or the median (accuracy of $\approx 0\%$). Figure~\ref{fig:3}C shows a comparison between the actual values of the order parameter $p$ as a function of the reduced temperature $T_r$ in a particular simulation and the values predicted by the $k$-nearest neighbors algorithm (with $k=15$). We note that the predictions are very close to the actual values of the order parameter, which confirms the efficiency and usefulness of our approach.

\subsection{Nematic textures of experimental samples}\label{res:experimental}

\begin{figure}[!ht]
\centering
\includegraphics[width=1\linewidth]{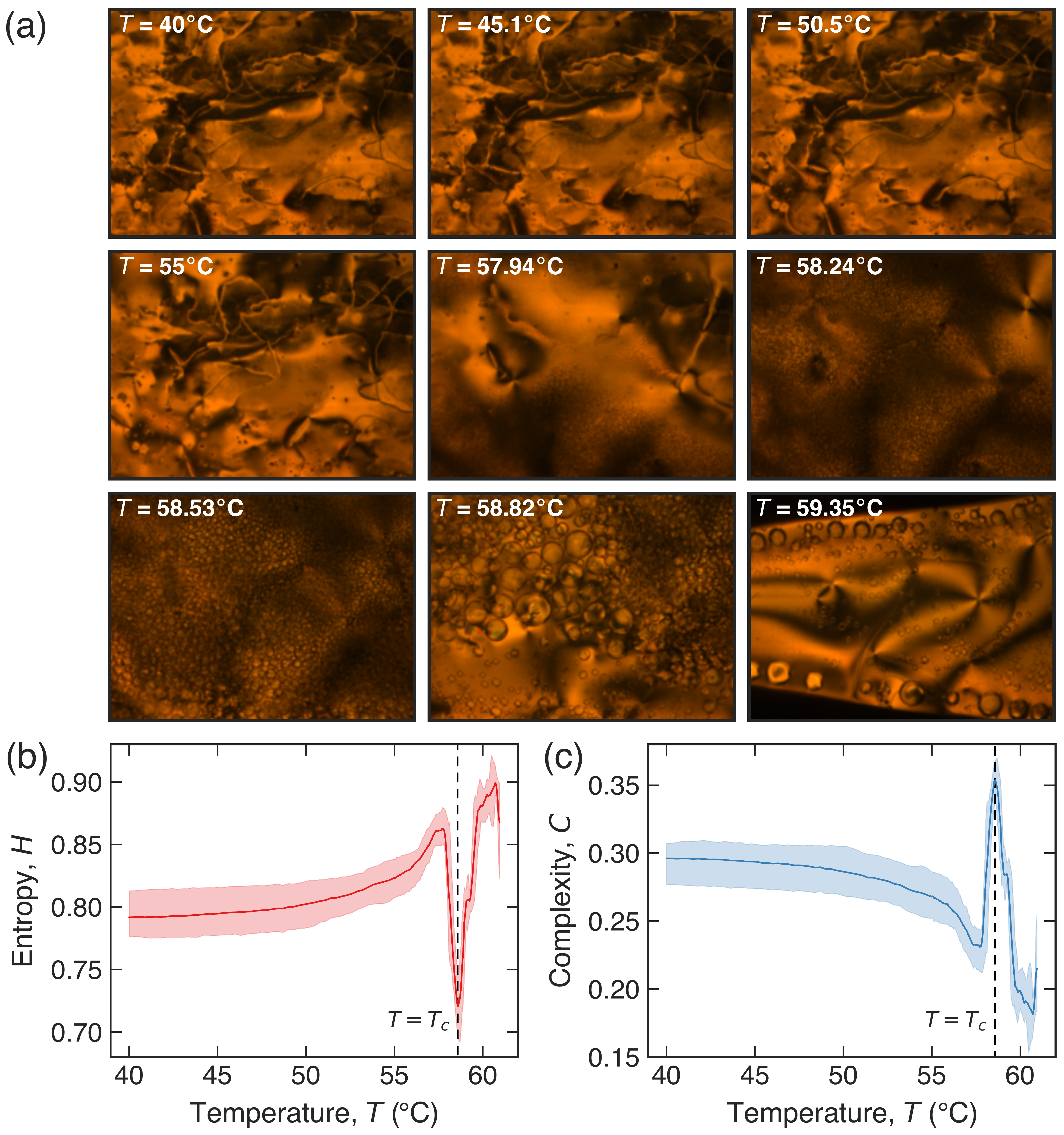}
\caption{{Examples of E7 liquid crystal textures at different temperatures and phases, and the dependence of the complexity measures on the temperature.} (a) Experimental textures of an E7 liquid crystal sample at different temperatures. We observe that practically no visual change is observed until the critical temperature $T_c\approx58~^\circ$C. (b) Dependence of the permutation entropy $H$ and (c) the statistical complexity $C$ on the temperature $T$. The solid curves represent the average of the quantities calculated over the results of six samples. The shaded areas are $95\%$ bootstrap confidence intervals. The vertical dashed lines indicate the critical temperature $T_c$. We note that the phase transition is properly identified by the extreme values of the complexity measures.}
\label{fig:4}
\end{figure}

In another application, we study nematic textures obtained from samples of the E7 liquid crystal, a multicomponent mixture composed by cyanobiphenyl and cyanoterphenol that is commonly employed in the industry for producing displays. This liquid crystal exhibits a nematic-isotropic transition at $T_c\approx58~^\circ$C~\cite{prl}. Details about the experimental procedures are provided in Appendix~\ref{app:experimental}, but it basically consists in using polarized optical microscope imaging for taking pictures of these textures at different temperatures $T$. Figure~\ref{fig:4}A shows examples of textures obtained from a sample with temperature varying from 40~$^\circ$C to 60~$^\circ$C. We observe that there are no visually significant changes in the patterns of these textures when the temperature increases until 55~$^\circ$C. As the sample temperature exceeds the critical temperature $T_c$, we observe the growth of isotropic domains which makes easier the visual distinction among these textures.

We collect data from six different E7 samples by following the same experimental protocol. Figures~\ref{fig:4}B and~\ref{fig:4}C depict the average behavior of the entropy $H$ and complexity $C$ on the temperature $T$. These results are similar to those obtained via Monte Carlo simulation (see Figure~\ref{fig:2}A), that is, $H$ tends to increase with $T$ and shows a sharp minimum at the critical temperature, while $C$ has a decreasing trend with $T$ and displays a maximum at the critical temperature. Once again, we observe that the values of $H$ and $C$ are well-defined functions of the temperature $T$ and capable of precisely identifying the nematic-isotropic transition.

\begin{figure}[!ht]
\centering
\includegraphics[width=1\linewidth]{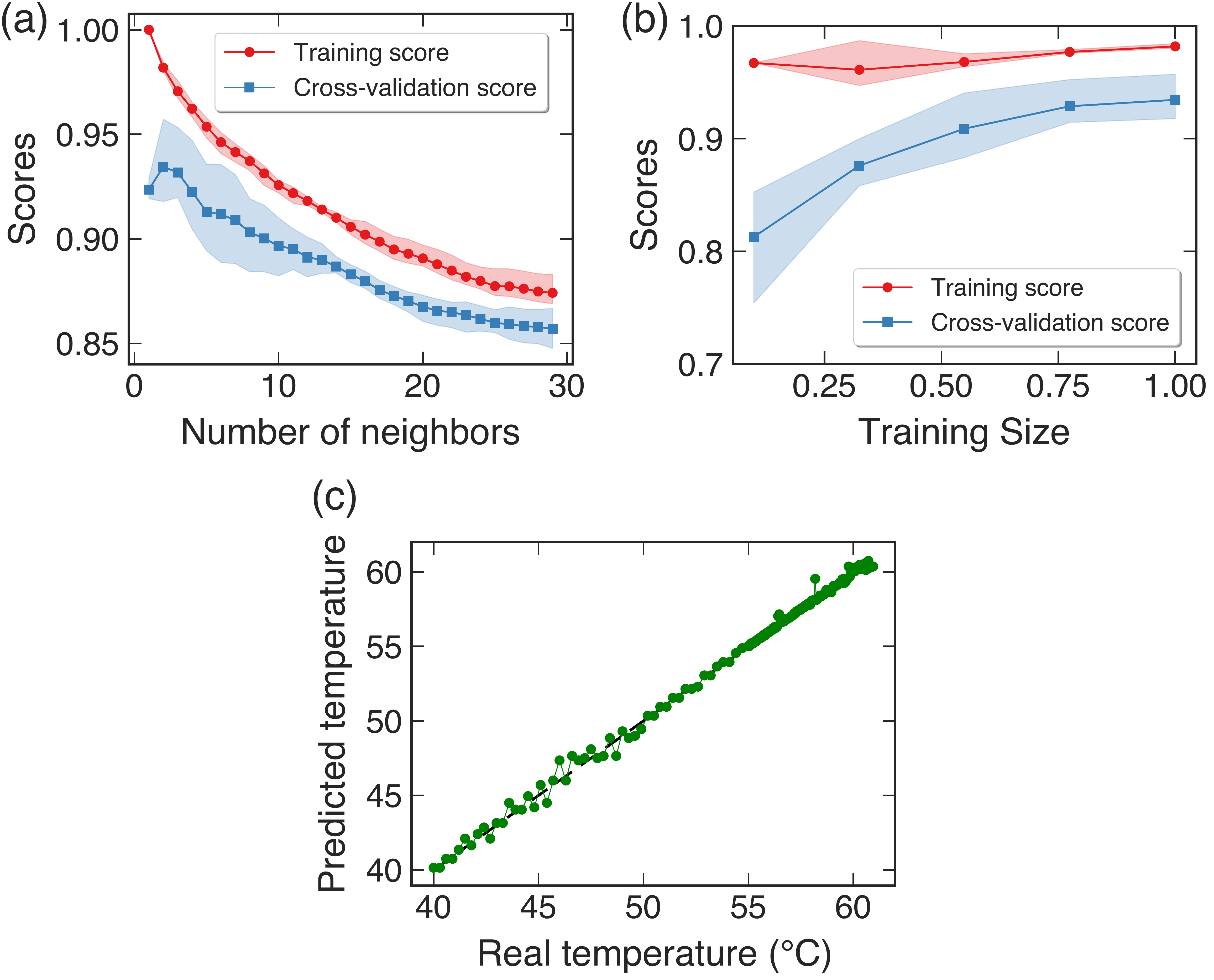}
\caption{{Predicting the temperature with a machine learning algorithm.} (a) Training and cross-validation scores of the $k$-nearest neighbors algorithm as a function of the number of neighbors $k$. The scores represent the coefficient of determination ($R^2$) of the relationship between the predicted and true values, and can be interpreted as the percentage (or fraction) of data variation that is explained by the model. Note that for $k=1$ the algorithm overfits the data, while for $k>3$ the model starts to underfit the data. (b) Learning curves, that is, the training and cross-validation scores as a function of the training size (fraction of the whole data) for $k=2$. The shaded areas in both plots represent the $95\%$ confidence intervals obtained in a 3-fold-cross-validation splitting strategy. We observe practically no significant improvement in the cross-validation score when more than $80\%$ of the data is used to train the model. (c) True versus predicted temperatures obtained by exposing the trained regressor to a set of experimental textures never presented before to the algorithm. The dashed line represents the 1:1 relationship. We observe an excellent agreement between the true and predicted temperatures, reinforcing the great accuracy achieved by the method ($\approx 93\%$).}
\label{fig:5}
\end{figure}

Differently from our previous results on the simulated textures, we now propose to predict the sample temperature $T$ (instead of the order parameter $p$) directly from the experimental textures. However, it is worth mentioning that in this simple experimental setup the order parameter can be estimated from the temperature~\cite{Wu2006}, so that predicting the temperature is comparable to predicting the order parameter. We also note that most of the models for liquid crystals fail to predict the order parameter for temperatures much lower than $T_C$~\cite{Wu2006}, and that in general, measuring the order parameter requires complicated procedures, such as light scattering experiments~\cite{PhysRevLett.25.503}. In order to perform these predictions, we proceed as in the simulated case, that is, we train a $k$-nearest neighbors algorithm for the regression task of predicting the values of $T$ from the image quantifiers ($H$ and $C$) and from a dummy variable that is zero for $T_c-0.05T_c\leq T\leq T_c$, 1 for $T>T_c$, and -1 for $T<T_c-0.05T_c$. Figure~\ref{fig:5}A shows the validation curves, where we observe that the algorithm overfits the data for $k=1$, while it starts to underfit the data for $k$ greater than $3$. Figure~\ref{fig:5}B depicts the learning curves for $k=2$, where we note no significant improvement in the cross-validation score when the training set exceeds 80\% of the whole data. Thus, the $k$-nearest neighbors algorithm achieve an accuracy of $\approx 93\%$ with $k=2$ and by using 80\% of data as training set. These scores are reduced in $\approx6\%$ when using the values of $H$ and $C$ separately, which reinforce the fact these complexity measures are not related to each other in a trivial manner. Furthermore, these scores are much higher than those obtained from regressors that always predict the expected values or the median (accuracy of $\approx 0\%$). Figure~\ref{fig:5}C illustrates the accuracy of these predictions by showing a scatter plot between the predicted temperature values and the actual values. We note that this relationship is an almost perfect 1:1 relation (indicated by the dashed line), which reinforces the overall quality of the $k$-nearest neighbors predictions and demonstrates the potential of our approach with experimental data.

\subsection{Simulated cholesteric textures}\label{res:cholesteric}

As a last application, we investigate simulated textures of cholesteric liquid crystals. These materials display a helical structure composed of layers in between which the preferential director axis varies periodically with a period (that is, the distance to complete a full rotation of the director axis) known as the pitch $\eta$. Among other properties, the pitch of a cholesteric liquid crystal defines the wavelength of the reflected light as a consequence of the Bragg reflection in short pitch materials~\cite{degennes}. The pitch length modifies the textures of these materials, so that, a cholesteric texture can mimic a nematic one for large values of the pitch or be entirely different for short pitches. In a cell treated to impose homeotropic alignment, for example, the pitch length determines if the texture observed is homeotropic, fingerprint or focal-conic~\cite{Wu2006}.

\begin{figure}[!ht]
\centering
\includegraphics[width=1\linewidth]{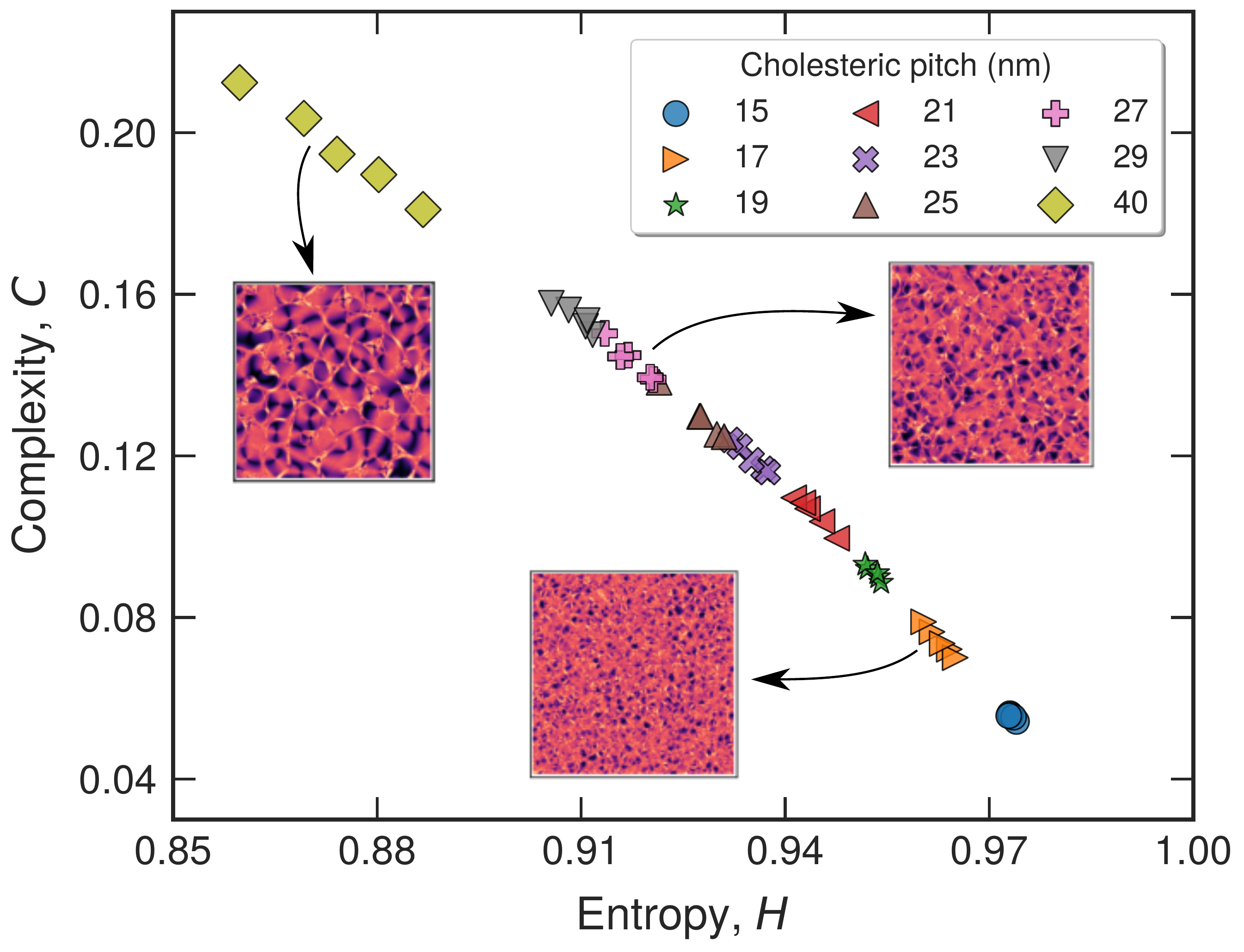}
\caption{{Discriminating among cholesteric textures with different pitches via the complexity-entropy plane.} Each colorful marker represents the values of $H$ and $C$ for 5 realizations of the cholesteric textures with different pitches (as indicated by the different markers). We note that the values of $H$ and $C$ for each cholesteric pitch are localized over a small region in the complexity-entropy plane. We further observe that the textures from small pitches are localized in a high entropy region, while those from large pitches are in a low entropy region. This result indicates that textures from large pitches are more ordered than those obtained for small pitches (as illustrated by the insets).}
\label{fig:6}
\end{figure}

Our goal in this case is to identify the pitch of a cholesteric liquid crystal based on the values of $H$ and $C$ obtained from the textures. To do so, we create a dataset of textures composed of one hundred replicas for each pitch value $\eta \in (15,17,19,21,23,25,27,29,40)$~nm, where $\eta=40$~nm is large enough to mimic a nematic texture. The optical textures are numerically obtained by solving the model described in Appendix~\ref{app:cholesteric}, which is based on the Landau-de Gennes theory~\cite{Ravnik2009}. In our simulations, we have used real values for the physical parameters of this model but a small lattice size (see Appendix~\ref{app:cholesteric} for further details). This choice leads to unrealistic cholesteric pitches but generates textures very similar to those obtained from experimental results. In particular, we use the process of quenching a cholesteric sample from the isotropic state to generate the set of textures. Figure~\ref{fig:6} shows the values of $H$ and $C$ estimated from five random selected textures for each value of the pitch as well as three typical textures for $\eta=17$~nm, $\eta=27$~nm and $\eta=40$~nm (insets of that figure). We observe that although there exists some overlapping, textures with different pitches tend to occupy different regions on the complexity-entropy plane, indicating that $H$ and $C$ are capable of distinguishing among different cholesteric textures. We further note that the larger the values of $\eta$, the higher the complexity and the lower the entropy values. Thus, large values of pitch produce textures more locally ordered, while small values generate textures that are locally more irregular.

\begin{figure}[!ht]
\centering
\includegraphics[width=1\linewidth]{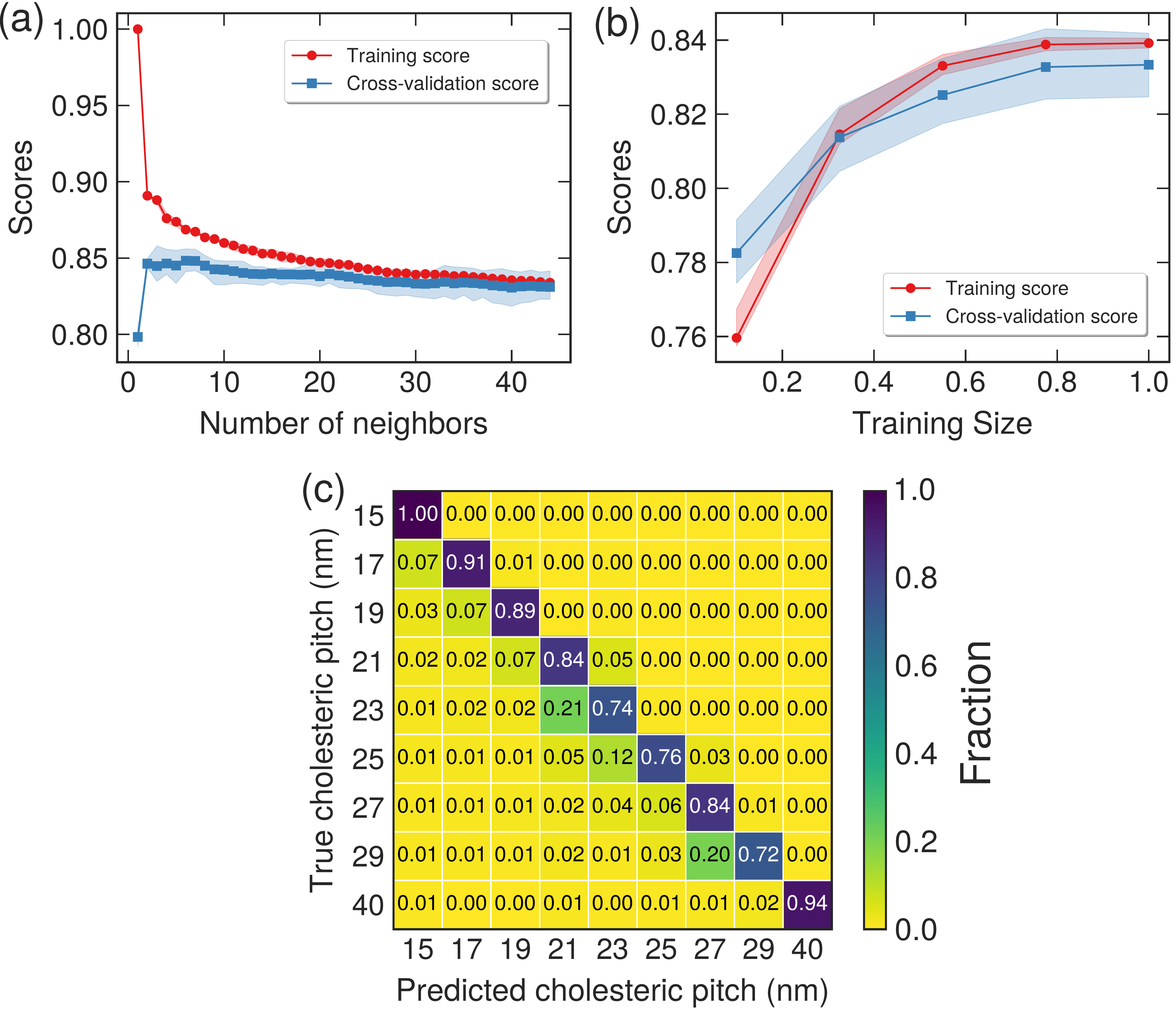}
\caption{{Predicting the cholesteric pitch with statistical learning algorithms.} Training and cross-validation scores (fraction of correct classifications) of the $k$-nearest neighbors algorithm as a function of the number of neighbors in (a) and as a function of the training size in (b). The shaded areas in both plots are the $95\%$ confidence intervals obtained in a 5-fold-cross-validation splitting strategy with the number of neighbors equal to 20. Note that for the number of neighbors smaller than 15 the algorithm overfits the data, but for more than 20 neighbors it starts underfitting the data. We also notice that when more than $60\%$ of the data is used to train the model (training size), there is no significant score improvement. (c) True and predicted cholesteric pitch. This confusion matrix shows the good performance achieved by the algorithm represented by the high values of right predictions in the diagonal. In some cases, the algorithm underestimates the pitch. This results from the overlap observed in the complexity-entropy plane in Figure~\ref{fig:6}.}
\label{fig:7}
\end{figure}

We train a $k$-nearest neighbors algorithm for the classification task of predicting the pitches solely based on the values of $H$ and $C$. Figure~\ref{fig:7}A shows the validation curves. We observe that this algorithm overfits the data when the number of neighbors is smaller than 3, and for larger number of neighbors, it slowly starts underfitting the data. We highlight that this simple algorithm achieves an accuracy of $\approx 85\%$, which is much higher than the baseline accuracy ($1/9\approx 11\%$) obtained from a classifier that makes uniformly random predictions. 
. The scores of $k$-nearest neighbors algorithm are reduced in $\approx 5\%$ when using the values of $H$ and $C$ separately, similarly to what happens with the E7 nematic textures. Also, the learning curves depicted in Figure~\ref{fig:7}B indicate that $\approx 60\%$ of the data is enough for fitting this algorithm to the cholesteric data. We have further estimated the confusion matrix, as shown in Figure~\ref{fig:7}C. The elements $f_{ij}$ of this matrix represent the fraction of textures with pitch $\eta_i$ that the algorithm predicts to have pitch $\eta_j$; thus, a perfect classifier is represented by an identity matrix ($f_{ij}=\delta_{ij}$). In practical applications, the closer to 1 are the diagonal elements of $f_{ij}$, the better is the performance of the classifier. In our case, we observe that nearly all nonzero elements of $f_{ij}$ are within a diagonal band of width 1 of this matrix, with the main diagonal concentrating at least $\approx72\%$ of the predictions. Thus, even when the algorithm incorrectly classifies the pitch of a texture (which occurs in about $15\%$ of the predictions), it tends to predict a pitch value that is very close to the actual value, with a higher probability of underestimate the value (notice that the elements preceding the main diagonal are larger than those appearing after). These results thus corroborate to the usefulness of our approach for investigating more complex textures.

\section{Conclusions}
We have proposed an approach for extracting physical properties of liquid crystals directly from textures images of these materials. Our method is based on estimating two simple complexity measures (permutation entropy and statistical complexity) directly from the textures, which are used as features in supervised learning tasks of regression and classification of physical parameters of these materials. We have demonstrated the usefulness and accuracy of this approach in a series of numerical and experimental applications. Our results have shown that the average order parameter can be directly estimated from images of nematic textures obtained from Monte Carlo simulations with accuracy of 98\%. Similar precision is obtained in the regression task of directly estimating the temperature from E7 liquid crystal textures at different temperatures and phases. We have further presented results based on cholesteric textures, in which we have probed the cholesteric pitch length with significant accuracy.

In spite of the significantly achieved accuracies, our approach is indeed quite simple and based on intuitive features very familiar to any physicist. Due to this underlying simplicity and also because our approach is very fast and scalable from the computational point of view, we believe it can be easily implemented and adapted for other more complex experimental situations involving the study of liquid crystals and perhaps for probing physical properties of different materials.

\acknowledgments
This research was supported by CNPq and CAPES. HVR thanks the financial support of the CNPq under Grants 440650/2014-3, 303642/2014-9, and 407690/2018-2. RSZ thanks the National Institute of Science and Technology Complex Fluids (INCT-FCx), and the Sao Paulo Research Foundation   (FAPESP – 2014/50983-3).

\appendix
\section{Complexity-entropy plane}\label{app:plane}
The normalized permutation entropy $H$~\cite{BandtPompe2002} and statistical complexity $C$~\cite{LopezManciniCalbet1995} are two complexity measures originally proposed for characterizing time series~\cite{RossoLarrondoMartinPlastinoFuentes2007}, and that were more recently generalized for considering higher dimensional data such as images~\cite{Ribeiro_etal2012,Zunino2016679}. We refer the more detail-oriented reader to the previously-cited references, where a complete description of these techniques can be found. Here we shall present both approaches through an illustrative example. For that, the matrix
\begin{equation*}
A =
\begin{bmatrix}
4 & 1 & 3 \\
3 & 5 & 6 \\
8 & 0 & 2
\end{bmatrix}
\end{equation*}
represents an hypothetical texture of size $3\times3$. The elements of this matrix indicate the light intensity transmitted through the sample around a particular site. For the experimental textures, these elements are obtained by averaging the shades of red, green, and blue of the image files in the RGB ``color space'' (see Appendix~\ref{app:experimental} for further details). We thus define sliding sub-matrices of size $d_x=2$ by $d_y=2$ (the embedding dimensions) in the form
\begin{equation*}
A_i =
\begin{bmatrix}
a_0 & a_1\\
a_2 & a_3
\end{bmatrix},
\end{equation*}
which for this particular example are
\begin{equation*}
A_1\!=\!
\begin{bmatrix}
4 & 1\\
3 & 5
\end{bmatrix},~
A_2\!=\!
\begin{bmatrix}
1 & 3\\
5 & 6
\end{bmatrix},~
A_3\!=\!
\begin{bmatrix}
3 & 5\\
8 & 0
\end{bmatrix},~\text{and}~
A_4\!=\!
\begin{bmatrix}
5 & 6\\
0 & 2
\end{bmatrix}\!.
\end{equation*}
Next, we associate a sequence of symbols to each sub-matrix for representing the ordinal patterns of occurrence of their elements. In this case, $A_1$ is associated with $\Pi_1=(1,2,0,3)$ since $a_1<a_2<a_0<a_3$, where $\Pi_1$ is the permutation that sorts the elements of $A_1$ in ascending order (line by line). Similarly, $A_2$ is represented by $\Pi_2=(0,1,2,3)$, since its elements are already in ascending order; $A_3$ is described by $\Pi_3=(3,0,1,2)$, since $a_3<a_0<a_1<a_2$; finally, $A_4$ is associated with $\Pi_4=(2,3,0,1)$ because $a_2<a_3<a_0<a_1$. In case of draws, the occurrence order of the tied elements is kept. By calculating the relative frequency of occurrence of each permutation, we estimate the probability distribution $P=\{p_i;~i=1,\ldots,n\}$ for each one of the $n=(d_xd_y)!$ ordinal patterns. Among the $(d_xd_y)!=24$ possible ordinal patterns, only four permutations have appeared once in our example, and therefore, $P=\{1/4,1/4,1/4,1/4,0,\dots,0\}$ is the ordinal distribution associated with the matrix $A$.

The ordinal distribution $P=\{p_i;~i=1,\ldots,n\}$ is thus used for estimating $H$ and $C$. The permutation entropy $H$ is the normalized Shannon entropy of $P$, that is, 
\begin{equation}
H(P) = \frac{1}{\ln(n)} \sum_{i=0}^{n} p_i \ln (1/p_i)\,,
\end{equation}
where $\ln(n)$ corresponds to the maximum value of the Shannon entropy $S(P)=\sum_{i=0}^{n} p_i \ln (1/p_i)$, occurring when all permutations are equally likely to occur ($p_i=1/n$). While the statistical complexity is defined by
\begin{equation}
C(P)=\frac{D(P,U) H(P)}{D^*}\,,
\end{equation}
in which 
\begin{equation}
D(P,U) = S\left(\frac{P+U}{2}\right) - \frac{S(P)}{2} - \frac{S(U)}{2}
\end{equation}
is the Jensen-Shannon divergence between $P$ and the uniform distribution $U=\{u_i=1/n;~i=1,\dots,n\}$ and $D^*$ is a normalization constant [obtained by calculating $D(P,U)$ when $P=\{p_i=\delta_{1,i};~i=1,\ldots,n\}$].

The entropy $H$ is a measure of ``disorder'' in the occurrence order of the elements of $A$. Values of $H\approx1$ indicate that these elements appear in random order, while values of $H\approx0$ imply that they tend to appear in a certain order. On the other hand, the values of $C$ quantify the ``structural'' complexity present in the matrix $A$. For a given value of $H$, the complexity $C$ can assume values between a minimum and a maximum and provides important additional information about the correlational structure of $A$ that is not properly carried out by the values of $H$. Mainly for this reason, we have used the diagram of $C$ versus $H$ (the so-called complexity-entropy plane) as a discriminating tool for investigating the liquid crystal textures. This framework has been successfully used in several applications with time series~\cite{Jovanovic2016,PhysRevE.89.012905,Stosic20161136,ribeiro2017characterizing,ribeiro2012complexity} and image analysis~\cite{schlemmer2015quantifying,antonelli2017permutation,sigaki2018history,antonelli2018mammographic}. In addition to their simplicity and intuitive meaning, these complexity measures are very fast and scalable from the computational point of view. This approach has also only the embedding dimensions $d_x$ and $d_y$ as ``tuning parameters''. However, this choice is not completely arbitrary, and the condition $(d_x d_y)!\ll n_x n_y$ must hold in order to obtain a reliable estimation of $P$. We have used $d_x=d_y=2$ in the study of simulated nematic textures and $d_x=2$ and $d_y=3$ in all other applications due to the dimensions of the matrices associated with the textures. However, very similar results are obtained when considering $d_x=3$ and $d_y=2$, $d_x=2$ and $d_y=3$ or $d_x=d_y=2$ in all applications.

\section{Monte Carlo simulations}\label{app:montecarlo}
In order to obtain the nematic textures analyzed in Section~\ref{res:montecarlo}, we have simulated a system composed of headless spins located over the sites of a tridimensional cubic lattice of dimensions $N_x\times N_y\times N_z$ (with $N_x=N_y=100$ and $N_z=20$). These spins have directions represented by unit vectors $\vec{u}_i$ [$i=(1,2,\dots,N)$, with $N=N_x N_y N_z=200,000$]. The spins in the first layer of the $z$-direction are fixed and point to the $y$-direction, while those in the last layer are fixed along the $x$-direction. These two layers of fixed spins mimic the surface region (denoted by $\mathcal{S}$) and supply an anchoring direction that twists the alignment of the spins across the sample. The other spins in the bulk region (denoted by $\mathcal{B}$) interact with their nearest neighbors via the Lebwohl-Lasher potential~\cite{lebwohl1972nematic} with periodic boundary conditions along the $x$ and $y$ directions. The Hamiltonian of this system can be written as
\begin{equation}
U_{\rm N}=\frac{1}{2} \sum_{\substack{i,j~\in~{\cal B} \\ i\neq j}}\Phi_{ij}+J\sum_{\substack{i~\in~{\cal B} \\ j~\in~{\cal S}}}\Phi_{ij},
\end{equation}
in which $J$ is the strength of the anchoring energy and
\begin{equation}
\Phi_{ij}=-\epsilon_{ij}\left(\frac{3}{2}\cos(\vec{u}_i \cdot \vec{u}_j)-\frac{1}{2}\right),
\end{equation}
with $\epsilon_{ij}=\epsilon$ when $i$ and $j$ are nearest neighbors, and zero otherwise. 

All textures are obtained with $J=1$, and the bulk spins are initially aligned making an angle with respect to the $x$-direction [$u_i=(\cos(0.3),\sin(0.3),0)$] for avoiding the appearance of unstable defects~\cite{chiccoli2015effect,chiccoli2017computer}. The lattice updates are obtained through the Metropolis algorithm. Each new configuration is generated by following the Barker-Watts technique~\cite{barker1969structure} and is accepted with probability $\exp(-\frac{\Delta U}{k_B T})$, where $\Delta U$ is the energy difference between the old and new states, $T$ the temperature, and $k_B$ the Boltzmann constant. A Monte Carlo step is completed when all spins are updated on average. 

The simulations start with a given reduced temperature $T_r=k_B T/\epsilon$, and the system is initially simulated for $10^{4}$ Monte Carlo steps to avoid transient behaviors. Next, we consider another $10^{4}$ Monte Carlo steps for estimating the average order parameter over each layer of the system. The local order parameter is the largest eigenvalue of the order matrix
\begin{equation}\label{eq:order_parameter}
Q_{ab}=\left\langle u_{i}^{(a)} u_{i}^{(b)}-\delta_{ab}\right\rangle,
\end{equation}
where $u_{i}^{(a)}$ and $u_{i}^{(b)}$ are the $a$-th and $b$-th components of the unitary vector $\vec{u}_i$ associated with the $i$-th spin, and $\delta_{ab}$ stands for the Kronecker delta. The order parameter across the entire sample ($p$) is calculated by averaging the mean value of each free layer of the system. This model is well-known to present a bulk phase transition at the critical temperature $T_c=1.1232$~\cite{fabbri1986monte}, that decreases when considering a confined sample with hybrid boundary conditions~\cite{chiccoli2003structures}. In our case, we have found that the critical temperature is $T_c=1.1075$. For simulations with reduced temperatures closer to this critical temperature, the system is initially simulated for $9\times10^{4}$ Monte Carlos steps for avoiding transient behaviors related to the phase transition. The textures are obtained by averaging the latest 50 Monte Carlo steps via the Stokes-Muller methodology~\cite{berggren1994computer}. This procedure consists in treating incoming light (parallel to the $z$-direction) as a Stokes vector, and describing each site as a Muller matrix. We have considered $n_e=1.66$ for the extraordinary refraction index, $n_0=1.50$ for the ordinary refraction index, a sample thickness of 5.3~$\mu$m, and wavelength of 545~nm for the incoming light. The resulting texture is represented by a matrix with dimensions $100 \times 100$, whose elements stand for the transmitted light intensity around a particular site of the surface area.

\section{Experimental proceedings and image files processing}\label{app:experimental}
The experimental textures analyzed in Section~\ref{res:experimental} are obtained via polarized optical microscope imaging of liquid crystal samples at different temperatures. The samples consist of rectangular capillaries with no surface treatment (300~$\mu$m$~\times~4$~mm) filled with the E7 mixture at $70~^{\circ}$C to avoid flow alignment. Next, the samples are cooled up to room temperature and placed on a temperature controller under the polarized optical microscope setup. We start taking pictures of textures from samples at $40~^{\circ}$C. The samples are slowly heated at a constant rate of $0.2~^\circ$C per minute, and pictures are taken every $90$~s until the temperature reaches $55^\circ$C. For higher temperatures, the heating rate is reduced to $0.05~^{\circ}$C per minute, and pictures are taken every $60$~s until the temperature of $61~^{\circ}$C is achieved. 

All acquired image files are in PNG format with dimensions of $2047$ pixels width by $1532$ pixels height, and 24 bits per pixel (8 bits for each one of the three layers in the RGB ``color space''). This means that each pixel has 256 possible intensities of red, green, and blue colors, allowing more than 16 million color variations. These files can be represented by a three-layer matrix of dimensions $n_x$ (image width) by $n_y$ (image height), in which each layer corresponds to a color channel whose elements are the color intensities (ranging from 0 to 255). We have calculated $0.2125 R + 0.7154 G + 0.0721 B$, that is, a weighted-average over the three layers, where $R$, $G$ and $B$ stand for the shade intensities of red, green, and blue colors of each pixel. This procedure corresponds to the gray-scale luminance (or reflectance) transformation~\cite{scikit-image}, which is considered to mimic the color sensibility of the human eye. This procedure yields a single matrix for each image file from which the values of $H$ and $C$ are calculated.

\section{Simulations of the continuum elastic theory}\label{app:cholesteric}

The cholesteric textures investigated in Section~\ref{res:cholesteric} are obtained via the continuum elastic theory. In particular, we have used the Landau-de Gennes approach~\cite{Ravnik2009} for describing the energy density $F$ associated with variations in the tensorial order parameter ${Q}$ around the equilibrium state. By letting $x_1$, $x_2$, and $x_3$ represent the spatial coordinates, the energy density can be written as
\begin{equation}
\begin{split}
  F&=\dfrac{L_1}{2} \dfrac{ \partial Q_{ij} }{\partial x_k} \dfrac{ \partial Q_{ij} }{\partial x_k} + \dfrac{L_2}{2} \dfrac{ \partial Q_{ij} }{\partial x_j} \dfrac{ \partial Q_{ik} }{\partial x_k} \\ 
   &+\dfrac{L_3}{2} Q_{ij} \dfrac{ \partial Q_{kl} }{\partial x_i} \dfrac{ \partial Q_{kl} }{\partial x_j} + \dfrac{4 \pi}{\eta}L_q \epsilon_{ikl}  Q_{ij} \dfrac{ \partial Q_{lj} }{\partial x_k}\\ 
  &+ \dfrac{A}{2} Q_{ij} Q_{ji} + \dfrac{B}{3} Q_{ij} Q_{jk} Q_{ki} \\
  &+\dfrac{C}{4} Q_{ij} Q_{jk} Q_{kl} Q_{li}\,,
\end{split}
\end{equation}
where $L_1$, $L_2$, $L_3$, and $L_q$ are elastic constants, $A$, $B$, and $C$ are thermodynamic parameters, and $\eta$ is the cholesteric pitch length. Here, we have assumed implicit summation in repeated indexes. The time evolution of the components of $Q_{ij}$ is given by
\begin{equation}\label{eq:TimeEvolution}
  \Gamma \dfrac{\partial Q_{ij}}{\partial t}= \left(\dfrac{\partial F({Q})}{\partial Q_{ij}}-\dfrac{d}{d x_k}\dfrac{\partial F({Q})}{\partial Q_{ij,k}}\right)\,,
\end{equation}
where $\Gamma$ is the liquid crystal rotational viscosity, $t$ is the time and $Q_{ij,k}$ is the derivative of $Q_{ij}$ relative to $x_k$. 

The system of equations~\ref{eq:TimeEvolution} is numerically solved via finite differences method in a uniform grid with $200\times200\times20$ grid points. All distance units are normalized by the grid distance $\delta x_1 =1$~nm, $\delta x_2 =1$~nm and $\delta x_3 =1$~nm. The liquid crystal parameters used are $A=-0.348$~MJ/(Km$^3$), $B=-2.133$~MJ/m$^3$, $C=1.733$~MJ/m$^3$, $\Gamma=0.3$~Pa s, $L_1=2.6$~pN, $L_2=2.6$~pN, $L_3=0.76$~pN, and $L_q=1.86$~pN. These parameters are obtained from the literature and are well-known to roughly describe the 5CB liquid crystal~\cite{degennes}. Thus, instead of rescaling the elastic and thermodynamic constants to unrealistic values~\cite{Ravnik2009}, we have used real values for these parameters but a small lattice size ($\delta x_i$). This choice implies in using small and unrealistic values for the cholesteric pitches. However, the textures produced with this approach are qualitatively very similar to those obtained when using unrealistic values for the physical parameters. Also, these textures are well-known to mimic well those obtained from experimental studies. We have considered different values for the pitch $\eta$ and the Dormand-Prince of fourth-order~\cite{dormand1980family} for the time integration. The initial condition is randomly chosen from a uniform distribution and periodic boundary conditions are considered in all directions for avoiding surface effects. Finally, the resulting optical textures are generated by applying the Jones $2\times2$ method~\cite{Wu2006} with the liquid crystal cells placed between crossed polarizers. This approach is the same used in reference~\cite{Sec2012} for generating the optical images.

\section{Implementation of the machine learning algorithms}\label{app:machinelearning}
Machine learning tasks include classification and regression. Classification is the task of predicting a discrete class label by providing a correctly labeled training set of data, whereas in regression tasks the algorithm predicts a continuous quantity. Thus, the predictions of the order parameter from the simulated nematic textures and the predictions of the temperature from the experimental nematic textures represent a regression task. On the other hand, the predictions of the discrete set of cholesteric pitches represent a classification task. For both tasks, we have used the $k$-nearest neighbors algorithms as implemented in the Python module scikit-learn~\cite{scikit-learn}. This is one of the simplest machine learning algorithms that make predictions based on the classes (or values) of $k$-closest neighbors~\cite{hastie2013elements}. The only parameter of this algorithm is the number of nearest neighbors $k$.

The best value of $k$ is chosen to simultaneously minimize the bias and variance errors of the predictions (the so-called bias-variance tradeoff~~\cite{hastie2013elements}). Bias errors occurs when the statistical learning algorithm is not complex enough to capture the underlying structure of the data (underfitting). On the other hand, variance errors appear when the algorithm is too complex and starts to model even the noise in the training set, but fail in predicting future observations in unseen data (overfitting). There is thus a trade-off between minimizing the bias and variance errors concerning the algorithm complexity.

To address this question, we employ a resampling strategy known as $n$-fold cross-validation~\cite{hastie2013elements}. This approach consists in randomly splitting the data into $n$ subsamples of approximately equal size. One of the subsamples is separated for validating the algorithm, and the remaining $n-1$ are used for training the algorithm. This process is thus repeated $n$ times. The accuracy obtained from the training set is the training score, and the one obtained from the validation set is the cross-validation score. At the end of $n$ repetitions, we have $n$ estimates for the training and cross-validation scores, from which we calculate average values and confidence intervals. The plot of the training and cross-validation scores as a function of model parameters (in our case, the number of nearest neighbors $k$) is known as validation curves. In our regression tasks, the scores represent the coefficient of determination ($R^2$) of the relationship between the predicted and true values. When positive, the $R^2$ is also interpreted as the percentage (or fraction) of the response variable variation that is explained by the model. Also, in our classification task, the score (or accuracy) represents the fraction of correct classifications. An underfitting situation happens when both training and validation scores are low; whereas overfitting is represented by good training scores and poor validation scores. The best bias-variance tradeoff occurs when the highest values are obtained for both scores. We have also estimated the learning curves, which represent the dependence of the training and the cross-validation scores on the size of the training set. This is an important matter for statistical learning methods because small training sets may not be enough for correctly fitting the model, while unnecessary data may introduce noise to the model.  

We have further compared the performance of the $k$-nearest neighbors algorithm with other more complex machine learning methods (namely: random forest, support vector machine, and neural network). However, the scores for these methods are very similar to those obtained for the $k$-nearest neighbors. For this reason and also due to the simplicity of the $k$-nearest neighbors, we have only reported the results for this machine learning method.

\bibliography{lc_textures_complexity_entropy}

\end{document}